\documentclass[11pt,a4paper]{article}

\usepackage[margin=2.5cm]{geometry}
\usepackage[utf8]{inputenc}
\usepackage[T1]{fontenc}
\usepackage{times}
\usepackage{setspace}
\usepackage{graphicx}
\usepackage{float}
\usepackage{booktabs}
\usepackage{longtable}
\usepackage{xcolor}
\usepackage{hyperref}
\hypersetup{colorlinks=true,linkcolor=blue,urlcolor=blue,citecolor=blue}
\usepackage{amsmath}
\usepackage{caption}

\usepackage{fancyhdr}
\title{A molecular clock for writing systems reveals the quantitative impact of\\imperial power on cultural evolution}
\author{Hiroki Fukui, M.D., Ph.D.\\[0.5em]
\small ORCID: 0009-0008-7122-522X\\[0.3em]
\small Research Institute of Criminal Psychiatry / Sex Offender Medical Center\\
\small Dept.\ Neuropsychiatry, Kyoto University\\[0.3em]
\small \href{mailto:fukui@somec.org}{fukui@somec.org}}
\date{April 2026}

\begin{document}

\maketitle

\section{Abstract}

Writing systems are cultural replicators whose evolution has never been studied quantitatively at global scale. We compile the Global Script Database (GSD): 300 writing and notation systems, 50 binary structural characters, and 259 phylogenetic edges spanning 5,400 years. Applying four methods --- phenetics, cladistics, Bayesian inference, and neural network clustering --- we find that scripts exhibit a detectable molecular clock. The best-fitting model (Mk+$\Gamma$ strict clock, incorporating Gamma-distributed rate variation across characters) yields a substitution rate of q = 0.226 substitutions/character/millennium (95\% CI: 0.034--1.22; $\Delta\text{BIC}$ = $-$4.1 versus relaxed clock; $\Delta\text{BIC}$ = $-$1,364.7 versus Mk without rate variation). The shape parameter $\alpha = 0.51$ reveals extreme rate heterogeneity among characters: physical features (directionality, writing medium, stroke construction) change at rates below 0.05/millennium, while social and structural features (pictographic origin, modular composition, user class) exceed 1.0/millennium --- a pattern consistent with functional constraint driving clock-like regularity.

Political interventions break this clock: deviation from expected divergence times correlates with intervention intensity (Spearman $\rho = 0.556,$ $p < 10^{-4}$), and this association holds when stimulus diffusion events are excluded ($\rho = 0.257$, $p = 5.9 \times 10^{-4}$) and when 9 features conceptually linked to political context are removed ($\rho = 0.505$, $p < 0.0001$). Per-character rate analysis reveals that intervention does not merely accelerate change but selectively rewrites deep structural features. The rate profiles of naturally transmitted and politically mediated scripts are only weakly correlated (Spearman $\rho = 0.320$), with the largest divergences concentrated in features defining script architecture --- pictographic origin, modular composition, and logographic component --- while physical substrate features resist transformation even under political coercion.

We identify 30 major script replacement events and rank their destructive impact using a multi-dimensional score validated by sensitivity analysis (Spearman $\rho = 0.989$ across 1,000 weight configurations). A ceiling effect suppresses independent invention wherever writing already exists (Fisher's exact OR = 0.054, $p < 10^{-6}$), and colonial contact is a powerful independent predictor of script extinction (Cox HR = 5.25, p = 0.0006). The Spanish Empire extinguished the most scripts (6 of 12 contacted, 50\% extinction rate), followed by the Empire of Japan (3 of 9, 33.3\%); the Spanish destruction of Maya is qualitatively unique as it eliminated one of only four independent inventions of writing. Feature coding was performed with LLM assistance (Claude, Anthropic) and validated by inter-rater reliability testing on a stratified sample of 40 scripts with two independent human coders blind to script dates and political context (Cohen's $\kappa = 0.877$; human-LLM $\kappa = 0.929$; Fleiss' $\kappa = 0.911$ across all three raters).

\newpage

\section{1. Introduction}

Writing is the most consequential information technology in human history. It enabled the accumulation of knowledge across generations, the administration of complex states, and the emergence of science, law, and literature as cumulative enterprises. Yet compared to spoken language --- whose evolution has been studied quantitatively for decades (Lieberman et al. 2007; Gray and Atkinson 2003; Greenhill et al. 2017; Pagel et al. 2007) --- the evolution of writing systems has received remarkably little computational attention. We know far more about how the word \textit{went} regularized to past tense than about how the letter A rotated from an ox head to its present form.

This asymmetry is not accidental. Writing systems present difficulties that spoken languages do not. They are consciously designed, subject to deliberate reform, and can be imposed or abolished by political decree. A language evolves by the accumulated, uncoordinated choices of millions of speakers; a script can be transformed overnight by the signature of a head of state. Atatürk replaced the Arabic script with Latin for Turkish in 1928. The Soviet Union Cyrillicized the writing systems of Central Asia within a single decade. The Spanish Empire destroyed Maya script --- one of only four independent inventions of writing in human history --- in the course of colonial conquest. These are not gradual evolutionary processes. They are interventions, and any model of script evolution must account for them.

The study of writing systems has traditionally been the domain of grammatology and epigraphy, disciplines that produce careful descriptive accounts of individual scripts but rarely attempt cross-system quantitative comparison. The major reference work, Daniels and Bright's \textit{The World's Writing Systems} (1996), is organized as a compendium of individual script descriptions rather than as a comparative framework. Computational approaches have been rare. Hosszú (2024) applied phenetic analysis to four Rovash (Old Hungarian-related) scripts and coined the term "Scriptinformatics," but the dataset of four scripts is too small for phylogenetic inference or statistical testing. We adopt the term "Computational Grammatology" to describe the approach taken here, distinguishing it from Hosszú's Scriptinformatics in three respects: the scale of the dataset (300 versus 4 systems), the application of phylogenetic and Bayesian methods (versus phenetics alone), and the explicit modeling of political intervention as a variable in script evolution. Phylogenetic methods have been applied to diverse cultural domains --- kinship systems, material culture, subsistence strategies --- but writing systems have been conspicuously absent from this program (Mace and Holden 2005), largely because no comparative dataset existed. Bentz and Dutkiewicz (2026) used entropy analysis to demonstrate that 40,000-year-old Paleolithic sign sequences have information density comparable to proto-cuneiform --- a striking finding that extends the timeline of symbolic notation but does not address the structural evolution of writing systems per se.

In biological systematics, the molecular clock hypothesis --- the idea that DNA sequences accumulate substitutions at a roughly constant rate --- has been one of the most productive tools for dating evolutionary divergences (Zuckerkandl and Pauling 1965). The clock has been applied to cultural data: Lieberman et al. (2007) showed that English irregular verbs regularize at a rate proportional to the inverse square root of their usage frequency, and Gray and Atkinson (2003) used Bayesian phylogenetics to date the divergence of Indo-European languages. But no one has tested whether writing systems exhibit clock-like evolution --- whether the structural features of scripts accumulate changes at a predictable rate over time.

This paper asks three questions:

\begin{enumerate}
  \item Do writing systems evolve at a clock-like rate --- and if so, what structural properties drive the regularity?
  \item Does political intervention break the clock, and does it alter the character of script change as well as its rate?
  \item Can we quantify the destructive impact of political power on writing systems?
\end{enumerate}

The answers, in brief: writing systems do exhibit a detectable molecular clock (q = 0.226 substitutions/character/millennium under the best-fitting Mk+$\Gamma$ model; strict clock favored, $\Delta\text{BIC}$ = $-$4.1). The clock's regularity arises from a hierarchy of feature-level change rates --- physical and technical features change slowly, constrained by motor habits and material substrates, while social and structural features change rapidly --- and the aggregate behavior is well described by a Gamma distribution with shape parameter $\alpha = 0.51.$ Political intervention breaks the clock (Spearman $\rho = 0.556,$ $p < 10^{-4}$), not merely by accelerating change, but by selectively altering which features change: the rate profiles of naturally transmitted and politically mediated scripts are only weakly correlated ($\rho = 0.320$), with the largest divergences concentrated in features defining a script's fundamental architecture. These three questions are addressed in turn in sections 3.2--3.3 (clock and its anatomy), 3.4 (political intervention), and 3.5--3.6 (ceiling effect and destruction). The destruction wrought by empires can be quantified and ranked. The Spanish Empire extinguished more scripts than any other polity (6 of 12 contacted, 50\% extinction rate), followed by the Empire of Japan (3 of 9, 33.3\%). The author is Japanese; that last finding is presented as self-critical scholarship, not as external accusation.

\newpage

\section{2. Materials and Methods}

\subsection{2.1 The Global Script Database}

We compiled the Global Script Database (GSD), a structured dataset of 300 writing and notation systems spanning approximately 5,400 years of documented writing history, from Proto-cuneiform (~3400 BCE) to scripts created in the 21st century. The database encompasses the full typological range of graphic communication systems: alphabets, abjads, abugidas, syllabaries, logosyllabic scripts, featural scripts, and pre-writing notation systems including pictographic, tally, token, and knot-record systems.

Each entry in the GSD contains 26 fields documenting the script's geographic origin (latitude, longitude, region, modern country), temporal range (start and end dates), typological classification (script type, phonetic encoding, symbol count, complexity level), functional characteristics (primary function, user class, whether it constitutes a full writing system), origin type (independent invention, adapted, stimulus diffusion, uncertain), political context, and bibliographic sources with confidence ratings.

Geographic coordinates and dates were assigned based on the earliest attested use of each script. For scripts with uncertain origins, we used the midpoint of the scholarly consensus range and flagged the entry with a reduced confidence rating. The database includes 158 living and 142 extinct systems.

Sources for the GSD include Daniels and Bright (1996), Robinson (2007), the ScriptSource database, the Unicode Consortium's script documentation, Rosa (2010, 2016) for Ryukyuan notation systems, and primary literature on individual scripts. Database construction, including initial feature coding, was performed with the assistance of Claude (Anthropic), a large language model, under human supervision. This methodological choice is discussed in detail in section 2.8.

The phylogeny connecting the 300 scripts consists of 259 parent-child edges, 14 scripts of uncertain phylogenetic placement (undeciphered or of unknown origin, listed separately), and 8 root nodes (4 independent inventions of writing and 4 independent notation system lineages). Each edge was classified by evidence level: 79 (30.5\%) are supported by major reference works (Daniels and Bright 1996; Coulmas 2003), 174 (67.2\%) reflect scholarly consensus in the specialist literature, and 6 (2.3\%) are inferred from structural similarity and historical context. The evidence classification for all 259 edges is provided in the supplementary data.

\subsection{2.2 Character Matrix}

A 300 $\times$ 50 binary character matrix was constructed to encode the structural features of each writing system. The 50 characters span the following domains:

\begin{enumerate}
  \item \textbf{Directionality} (4 characters): left-to-right, right-to-left, top-to-bottom, boustrophedon
  \item \textbf{Structural type} (3 characters): pictographic origin, phonetic component, logographic component
  \item \textbf{Graphemic properties} (10 characters): diacritics, independent vowel signs, consonant-inherent vowel (abugida property), determinatives, punctuation, stroke-based construction, curved strokes, angular strokes, glyph symmetry, ligature frequency
  \item \textbf{Layout} (2 characters): baseline alignment, modular composition (radical-based structure)
  \item \textbf{Glyph inventory size} (3 characters): under 50 glyphs, 50--200 glyphs, over 200 glyphs
  \item \textbf{Functional domain} (8 characters): administration, religion, commerce, literature, divination, personal communication, monumental inscription, everyday use
  \item \textbf{User class} (3 characters): elite/royal users, trained scribes, common population
  \item \textbf{Origin and transmission} (6 characters): state-sponsored, community-developed, individually invented, independent invention, adapted from parent, stimulus diffusion
  \item \textbf{Political context} (3 characters): imposed by colonizer, voluntarily borrowed, inherited within tradition
  \item \textbf{Writing medium} (6 characters): clay, stone, papyrus/paper, wood/bark/palm leaf, metal, digital
  \item \textbf{Writing tool} (2 characters): stylus/chisel, brush/pen
\end{enumerate}

Missing data constituted 2.7\% of the matrix. Missing values were coded as absent (0) for the purpose of distance calculations, following standard practice in morphological phylogenetics (Lewis 2001).

\subsection{2.3 Phylogenetic Reconstruction}

We applied four complementary methods to reconstruct the evolutionary relationships among writing systems:

\textbf{Phenetics.} Pairwise Sørensen-Dice distances were computed from the binary character matrix. Two clustering algorithms were applied: Weighted Pair Group Method with Arithmetic Mean (WPGMA) and Neighbor-Joining (NJ). Principal Component Analysis (PCA) on the distance matrix extracted 3 components explaining 52.5\% of total variance.

\textbf{Cladistics.} Maximum Parsimony analysis was performed using the Fitch algorithm with Subtree Pruning and Regrafting (SPR) heuristic search. Tree statistics: Consistency Index (CI) = 0.10, Retention Index (RI) = 0.85. The low CI reflects extensive homoplasy, expected in cultural data where horizontal transfer is common.

\textbf{Bayesian inference.} We implemented a Markov chain Monte Carlo (MCMC) analysis using the Mk model (Lewis 2001), the standard model for discrete morphological characters. The Mk model assumes that character state changes follow a continuous-time Markov process with a single rate parameter \textit{q}. We used the Felsenstein pruning algorithm (Felsenstein 1981) for likelihood computation. MCMC was run with 16 walkers for 10,000 iterations with 2,000 burn-in, implemented via the emcee Python package. Convergence was assessed via effective sample size (ESS > 100 for all parameters).

\textbf{Neural network.} An autoencoder architecture (50$\rightarrow$20$\rightarrow$8$\rightarrow$20$\rightarrow$50) was trained to learn a compressed representation of script features. The 8-dimensional latent space was visualized via UMAP projection (2D). The autoencoder achieved 42.6\% classification accuracy on script type prediction from the latent representation (versus 7\% expected by chance for 14 categories), indicating that the latent space captures meaningful typological structure. Unsupervised clustering on the latent space identified 10 cross-type clusters not predicted by traditional classification.

\subsection{2.4 Molecular Clock Analysis}

We tested whether writing systems evolve at a clock-like rate by applying molecular clock models adapted from biological systematics. Two substitution models were compared: the standard Mk model (Lewis 2001), which assumes a single rate parameter q across all characters, and the Mk+$\Gamma$ model, which allows rate variation among characters by drawing per-character rates from a Gamma distribution with shape parameter $\alpha$ (Yang 1994). The Gamma distribution was discretized into K = 4 rate categories following standard practice. When $\alpha$ is large, all characters evolve at similar rates and the model converges to Mk; when $\alpha$ is small, a few characters evolve rapidly while most change slowly.

For each substitution model, two clock models were tested. The strict clock assumes a single substitution rate q across all branches of the phylogeny:

$$d(i,j) = q \times r_k \times t(i,j)$$

where $d(i,j)$ is the character distance between parent script $i$ and child script $j$, $t(i,j)$ is the archaeological divergence time in millennia, and $r_k$ is the relative rate for character category $k$ (equal to 1 for all characters under the Mk model). The relaxed clock allows rate variation among branches via a log-normal distribution.

This yields four models: Mk strict, Mk relaxed, Mk+$\Gamma$ strict, and Mk+$\Gamma$ relaxed, compared via the Bayesian Information Criterion (BIC). Bayesian inference was implemented via Markov chain Monte Carlo (MCMC) using the emcee affine-invariant ensemble sampler (16 walkers, 10,000 iterations, 2,000 burn-in). Prior distributions: $q \sim \text{Exponential}(\text{mean}=1)$, $\text{scale} \sim \text{Exponential}(\text{mean}=10)$, $\alpha \sim \text{Exponential}(\text{mean}=1)$. Convergence was assessed via the Gelman-Rubin statistic ($\hat{R} < 1.01$ for all parameters across all four models), effective sample size (ESS > 2,400 for all parameters), and visual inspection of trace plots (Supplementary Figure S2).

To evaluate the clock's predictive power, we estimated divergence times for all 246 parent-child pairs with known archaeological dates and computed Pearson correlation, $R^2$, and mean absolute error (MAE) between estimated and archaeological dates.

\subsection{2.4.1 Per-Character Rate Estimation}

To investigate why writing systems exhibit clock-like behavior despite the diversity of their structural features, we estimated the substitution rate of each of the 50 characters individually. For each parent-child pair, we recorded whether each character changed state (0$\rightarrow$1 or 1$\rightarrow$0) and divided by the divergence time in millennia, yielding a per-character rate. Rates were averaged across all 246 pairs and ranked. Bootstrap 95\% confidence intervals were computed from 10,000 resamples.

To test whether political intervention changes the profile of which features are modified --- not merely the overall rate --- we partitioned the 246 pairs into natural transmission (n = 128) and political intervention (n = 113) groups and computed per-character rates separately for each. The similarity of the two rate profiles was assessed via Spearman rank correlation. Features with the highest intervention-to-natural rate ratio were identified as intervention-preferential.

\subsection{2.4.2 Robustness: Reduced Feature Matrix}

Nine of the 50 characters encode political context (3 characters: imposed by colonizer, voluntarily borrowed, inherited within tradition) or origin and transmission (6 characters: state-sponsored, community-developed, individually invented, independent invention, adapted from parent, stimulus diffusion). Because these characters partially overlap with the intervention classification used as the independent variable in the deviation analysis, we repeated the full molecular clock analysis using a reduced 41-character matrix with these 9 characters excluded. This tests whether the clock signal depends on features that are conceptually linked to the political processes under study.

\subsection{2.5 Intervention Classification and Deviation Scoring}

Each of the 300 scripts was classified according to the type of political intervention, if any, that shaped its transmission or adoption:

\begin{itemize}
  \item \textbf{None} (n = 156): natural transmission from parent to child script
  \item \textbf{Stimulus diffusion} (n = 73): creation of a new script inspired by exposure to the concept of writing, without direct copying of graphemes
  \item \textbf{Uncertain} (n = 27): ambiguous or mixed mechanisms
  \item \textbf{Religious conversion} (n = 16): script change driven by religious institutions
  \item \textbf{Reform} (n = 12): deliberate restructuring of an existing script by political authority
  \item \textbf{Imposed standard} (n = 9): enforcement of a standardized form across a population
  \item \textbf{Colonial imposition} (n = 7): replacement of indigenous scripts by colonial powers
\end{itemize}

These counts cover all 300 scripts in the GSD. Of these, 246 participate in at least one parent-child pair and are included in the molecular clock deviation analysis (section 3.4). The remaining 54 scripts --- root nodes, proto-writing systems, and scripts of uncertain phylogenetic placement --- lack a parent script against which divergence can be measured and are excluded from the clock analysis, though they are included in the ceiling effect and survival analyses.

Each script was additionally assigned an intervention intensity score on a 0--5 ordinal scale, where 0 represents purely natural transmission and 5 represents complete replacement of the prior script.

Deviation from the molecular clock was computed for each script as the absolute difference between the clock-estimated divergence time and the archaeological divergence time. We tested the association between deviation and intervention intensity using Spearman rank correlation. Differences in deviation across intervention types were tested using the Kruskal-Wallis H test with Dunn's post-hoc comparisons.

\subsection{2.6 Destruction Score}

For 30 identified script replacement events --- episodes in which a dominant script was replaced by another through political action --- we computed a multi-dimensional destruction score integrating five equally-weighted normalized components: (1) feature distance (Hamming) between old and new scripts, (2) transition speed (inverse of years for the transition to complete), (3) affected population size at the time of replacement, (4) survival status of the old script (binary: extinct = 1, survived = 0), and (5) phylogenetic distance between old and new scripts in the GSD tree.

Imperial script destruction counts were compiled by attributing each extinction event to the political entity responsible.

To assess robustness to the choice of equal weighting, we performed a sensitivity analysis drawing component weights from a Dirichlet distribution across 1,000 random configurations and computing the rank distribution of each event.

\subsection{2.6.1 Robustness Check: Stimulus Diffusion Exclusion}

Stimulus diffusion scripts are, by definition, created without direct graphemic inheritance from a parent script, making their "distance" from the parent partly tautological. To test whether the association between political intervention and clock deviation is driven by this category, we repeated the deviation analysis (section 3.4) with all stimulus diffusion scripts (n = 71) excluded. Of the 73 stimulus diffusion scripts in the GSD, 71 have parent-child pairs in the phylogeny; the remaining 2 are root nodes without computable divergence times and are excluded from the deviation analysis.

\subsection{2.6.2 Definition of Script Extinction}

A script is classified as "extinct" in the GSD if it has no living community of active users. The extinction date is assigned as the approximate date of last documented use, based on the most recent dated inscription, manuscript, or scholarly report. Scripts that persist only in ceremonial, artistic, or scholarly contexts (e.g., Sanskrit written in Devanagari) are classified as "living" if they remain in active use by a community, even if the associated spoken language is no longer in everyday use.

\subsection{2.6.3 MCMC Convergence Diagnostics}

The Bayesian analysis was implemented using the emcee affine-invariant ensemble sampler with 16 walkers and 10,000 iterations (2,000 burn-in). Convergence was assessed using three criteria. First, the Gelman-Rubin statistic ($\hat{R}$) was computed by splitting the 16 walkers into two groups of 8; all parameters yielded $\hat{R} < 1.01$ (strict clock: $\hat{R}_q = 1.003$, $\hat{R}_{\text{scale}} = 1.003$; relaxed clock: $\hat{R}_q = 1.007$, $\hat{R}_{\text{scale}} = 1.007$, $\hat{R}_{\text{rate\_var}} = 1.007$), well below the convergence threshold of 1.1. Second, effective sample sizes (ESS) were computed using integrated autocorrelation time estimates: strict clock $\text{ESS}_q = 3{,}419$, $\text{ESS}_{\text{scale}} = 3{,}428$; relaxed clock $\text{ESS}_q = 2{,}770$, $\text{ESS}_{\text{scale}} = 2{,}778$, $\text{ESS}_{\text{rate\_var}} = 2{,}935$. All ESS values exceed the recommended minimum of 400 for reliable posterior estimation. Third, the acceptance rate for the strict clock was 0.715, within the optimal range (0.2--0.5 for Metropolis-Hastings; affine-invariant samplers typically run higher). Trace plots for all parameters are provided in Supplementary Figure S2.

\subsection{2.7 Ceiling Effect Extension: Cox Model with Colonial Contact}

To assess whether the ceiling effect is independent of colonial contact, we extended the Cox proportional hazards model by adding colonial contact as a binary covariate. This variable is defined broadly: a script is coded as colonial contact = 1 if it was created, transmitted, or used under conditions of colonial rule, regardless of whether the colonial power directly imposed a replacement script (n = 31). This definition is broader than the intervention classification category 'colonial imposition' (n = 7), which is restricted to cases where a colonial power actively replaced an indigenous script. The broader definition captures scripts that experienced colonial pressure without direct script replacement and provides greater statistical power for the survival analysis. Three models were compared: the original model (ceiling + covariates), an extended model (ceiling + colonial contact + covariates), and an interaction model (ceiling $\times$ colonial contact + covariates). Models were compared via AIC.

\subsection{2.8 LLM Transparency and Limitations}

This study used Claude (Anthropic) as a computational tool for database construction, feature coding, statistical analysis (Python script generation), and draft writing. We report this explicitly because LLM-assisted research raises specific concerns that must be addressed:

\textbf{Feature coding circularity.} The character matrix was coded by an LLM whose training data includes information about the archaeological dates of writing systems. The molecular clock analysis then "predicts" those dates from feature distances. This creates a risk of circular reasoning: the clock may appear to work because the features were coded with implicit knowledge of the dates. We mitigate this concern through four lines of evidence: (a) the 50 structural characters are defined independently of chronology, (b) the clock fails systematically where political interventions distort the feature record, which would not occur if features were simply back-calculated from dates, (c) two independent human coders blind to script dates and political context coded a stratified sample of 40 scripts (spanning all intervention types, script types, and historical eras), achieving Cohen's $\kappa = 0.877$ with each other and $\kappa = 0.929$ against the LLM coding (Fleiss' $\kappa = 0.911$ across all three raters; n = 2,000 binary judgments), and (d) per-feature analysis identified only one problematic feature (curved\_strokes: $\kappa = -0.04$, a prevalence-driven paradox with 92.9\% raw agreement) while all domain-level $\kappa$ values exceeded 0.83.

\textbf{Database completeness.} The GSD contains 300 systems but is not exhaustive. African and Southeast Asian small-scale notation systems are likely underrepresented. The preprint-stage findings should be interpreted with this caveat.

\textbf{Reproducibility.} All data (GSD, character matrix, phylogeny, intervention classifications) and all Python analysis scripts are provided as supplementary materials.

\newpage

\section{3. Results}

\subsection{3.1 Script Phylogeny and the Structure of the Global Script Database}

The GSD contains 300 writing and notation systems connected by 259 phylogenetic edges, rooted at 8 independent nodes (4 independent inventions of writing and 4 independent notation system lineages), with 14 additional scripts of uncertain phylogenetic placement, including the four independently invented writing traditions (Sumerian cuneiform, Egyptian hieroglyphs, Chinese characters, Mesoamerican scripts) and two proto-writing lineages. The phylogeny reveals that 65\% of the world's writing systems serve language families other than the one spoken by the script's inventors --- writing, as a technology, travels independently of the languages it encodes.

PCA on the 300 $\times$ 50 character matrix yielded three principal components explaining 52.5\% of the total variance. The first component (22.1\%) separated phonetic scripts (alphabets, abugidas) from logographic and pictographic systems. The second component (17.3\%) distinguished cursive, connected scripts from block-based, discrete-character systems. The third component (13.1\%) captured the distinction between scripts with and without vowel notation.

The autoencoder's 8-dimensional latent space, projected to 2D via UMAP, revealed 10 cross-type clusters that challenge the traditional typological classification. The most salient finding is that the boundary between alphabets and abugidas --- treated as categorically distinct in standard grammatology --- is not sharp in feature space. Several South and Southeast Asian abugidas cluster with European alphabets rather than with other abugidas, suggesting that the alphabet-abugida distinction may be an artifact of classificatory convention rather than a structural discontinuity.

\begin{figure}[H]
\centering
\includegraphics[width=0.9\textwidth]{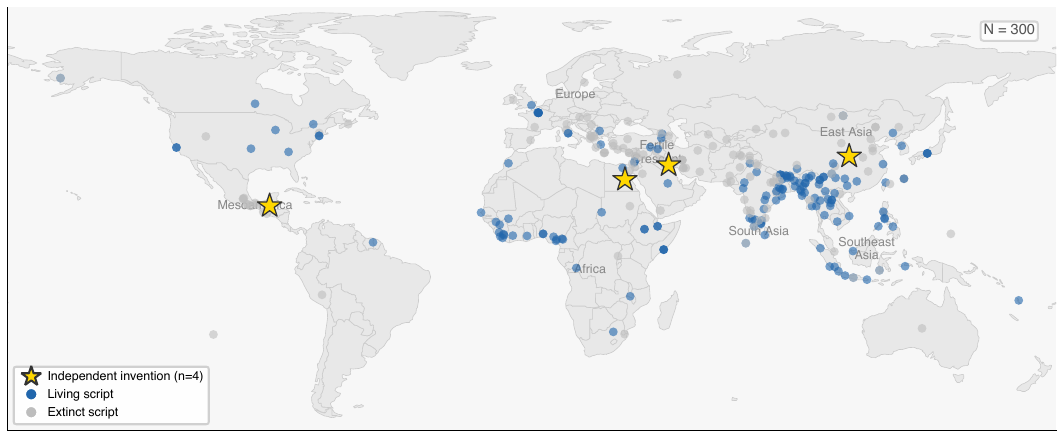}
\caption{Figure 1. Geographic distribution of 300 writing systems in the Global Script Database. Stars indicate independently invented scripts (n=4). Blue: living; gray: extinct.}
\label{fig:1}
\end{figure}

\begin{figure}[H]
\centering
\includegraphics[width=0.9\textwidth]{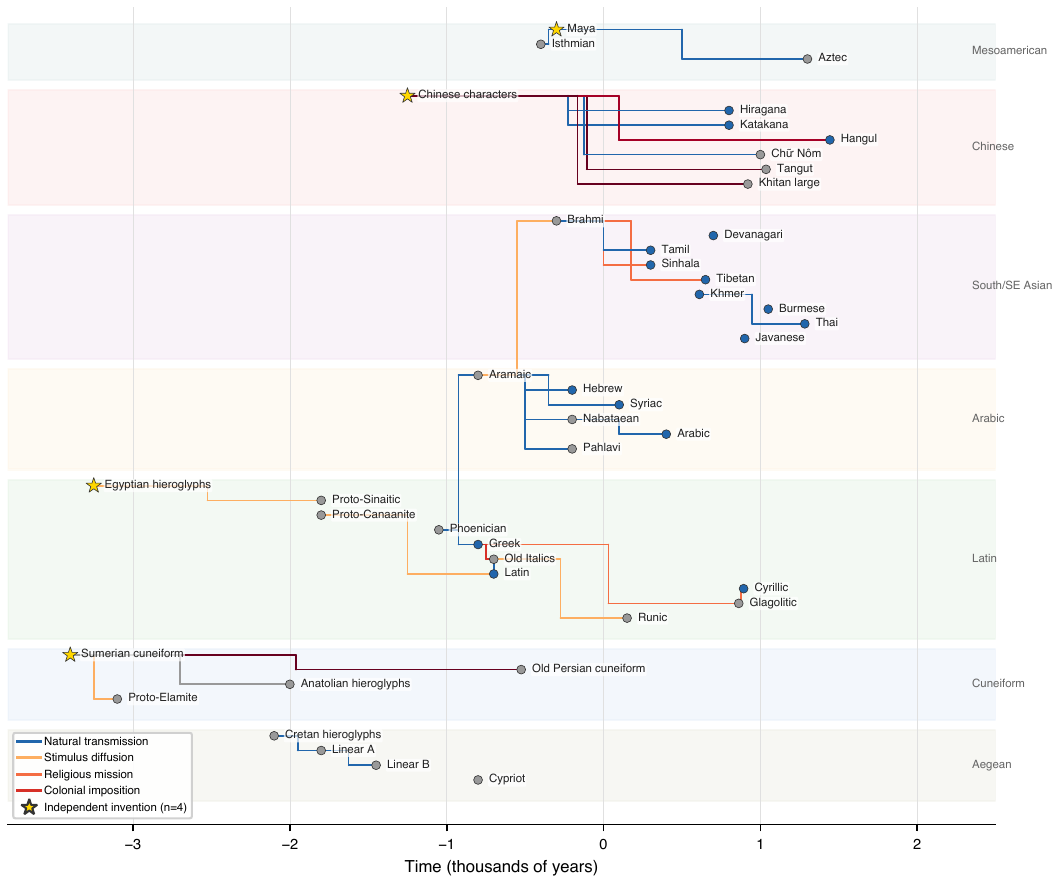}
\caption{Figure 2. Selected phylogenetic lineages of writing systems. Line color indicates transmission mechanism: blue = natural, orange = stimulus diffusion, salmon = religious mission, red = colonial imposition. Stars indicate independent inventions (n=4). Time axis in thousands of years.}
\label{fig:2}
\end{figure}

\subsection{3.2 The Molecular Clock of Writing}

Four models were compared: Mk strict clock, Mk relaxed clock, Mk+$\Gamma$ strict clock, and Mk+$\Gamma$ relaxed clock. The introduction of Gamma-distributed rate variation across characters (Mk+$\Gamma$) produced a dramatic improvement in model fit ($\Delta\text{BIC} = -1{,}364.7$ for Mk+$\Gamma$ strict versus Mk strict), confirming that the 50 structural characters evolve at substantially different rates. The estimated shape parameter $\alpha = 0.51$ (95\% CI: 0.46--0.58) indicates a strongly right-skewed rate distribution: most characters change slowly while a few evolve rapidly.

Within the Mk+$\Gamma$ framework, the strict clock was again favored over the relaxed clock ($\Delta\text{BIC} = -4.1$), indicating that a single branch-level substitution rate adequately describes writing system evolution even after accounting for character-level rate heterogeneity. The estimated substitution rate under the best-fitting model (Mk+$\Gamma$ strict) was $q = 0.226$ substitutions/character/millennium (95\% CI: 0.034--1.22). The upward shift from the Mk estimate ($q = 0.157$) reflects the correction of a downward bias introduced by the equal-rate assumption: when slowly evolving characters are given equal weight to rapidly evolving ones, the average rate is pulled toward the conservative end of the distribution.

The wide confidence interval warrants caution: it reflects both genuine uncertainty in the rate estimate and the heterogeneity inherent in cultural data. We return to the interpretation of this interval in Discussion 4.1.

Across all 246 parent-child pairs with known archaeological dates, the molecular clock showed strong overall correlation between estimated and archaeological divergence times ($r = 0.959$, $R^2 = 0.919$, MAE = 979 years; Figure 3a). When partitioned by intervention status, scripts transmitted through natural processes showed $R^2 = 0.876$ and MAE = 879 years (n = 128; Figure 3b), while scripts whose transmission involved political intervention showed $R^2 = 0.957$ and MAE = 1,110 years (n = 113; Figure 3c). Five parent-child pairs with uncertain intervention classification were excluded from the partition analysis, yielding n = 128 (natural) + n = 113 (intervention) = 241.

The higher $R^2$ in the intervention group is counterintuitive. It does not indicate greater clock precision; rather, it reflects the narrower temporal range of the intervention group (predominantly within the last 2,000 years), which mechanically inflates $R^2$. When the natural group is restricted to pairs diverging within the past 2,000 years (n = 89), its $R^2$ rises to 0.914, approaching the intervention group's value. MAE --- which is insensitive to range effects --- remains the more reliable indicator of clock deviation, and the 26\% increase in MAE for the intervention group (1,110 versus 879 years) constitutes the primary evidence that political intervention increases deviation from the clock.

\textbf{Robustness check: reduced feature matrix.} The full molecular clock analysis was repeated with the 9 political-context and origin-transmission characters excluded (41-character matrix). The strict clock remained strongly favored ($\Delta\text{BIC} = -15.2$), and the Spearman correlation between intervention intensity and clock deviation remained significant ($\rho = 0.505$, $p < 0.0001$). $R^2$ decreased from 0.919 to 0.843 and MAE increased from 979 to 1,539 years, confirming that the excluded features contributed to the clock's apparent precision. However, the qualitative conclusions --- the existence of a detectable clock, and the association between political intervention and clock deviation --- are robust to their exclusion. A finer decomposition confirms this pattern: removing only the 3 political-context characters (47-feature matrix) yields $R^2$ = 0.917 and $\rho = 0.556,$ virtually identical to the full 50-feature analysis; removing only the 6 origin-and-transmission characters (44-feature matrix) yields $R^2$ = 0.924 and MAE = 849 years, marginally improving on the full matrix. The precision loss in the 41-feature analysis is thus attributable to the cumulative removal of both groups, not to contamination by either group alone.

\begin{figure}[H]
\centering
\includegraphics[width=0.9\textwidth]{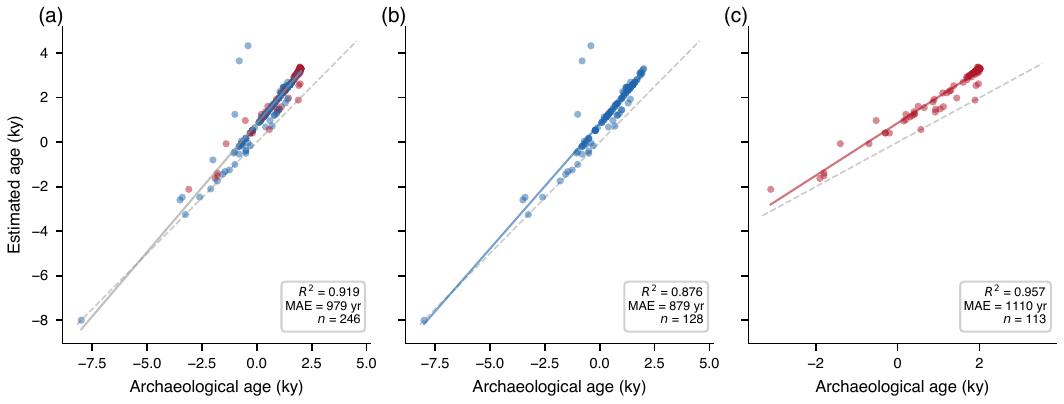}
\caption{Figure 3. Molecular clock of writing systems. (a) All 246 parent-child pairs ($R^2 = 0.919$, MAE = 979 yr). (b) Natural transmission only (n = 128, $R^2 = 0.876$, MAE = 879 yr). (c) Political intervention only (n = 113, $R^2 = 0.957$, MAE = 1,110 yr). Dashed line: y = x.}
\label{fig:3}
\end{figure}

\subsection{3.3 The Anatomy of Script Change}

Per-character substitution rates spanned more than three orders of magnitude, from stroke-based construction (0.001 changes/millennium) to pictographic origin (2.17 changes/millennium). The most conservative features were physical and technical: stroke-based construction, writing medium (clay, digital), top-to-bottom directionality, and use for divination. The most rapidly evolving features were social and structural: pictographic origin, modular composition, community-developed status, elite user class, and intermediate glyph inventory size (50--200 glyphs).

This hierarchy is consistent with a functional constraint hypothesis: features tied to the physical substrate of writing --- the direction of hand movement, the nature of the writing surface, the basic tooling of stroke production --- are constrained by motor habits, material properties, and embodied practice. Features describing the social positioning of a script or its high-level structural architecture are free to change as political and cultural contexts shift.

A Kruskal-Wallis test across the 11 feature domains yielded H = 12.49 (p = 0.052), a marginal result reflecting within-domain variance. However, the magnitude of the between-domain contrast is substantial: social-context features changed at a mean rate of 1.15/millennium, approximately six times faster than technical features (0.19/millennium).

The rate profiles of naturally transmitted and politically mediated scripts were only weakly correlated (Spearman $\rho = 0.320,$ p = 0.024), indicating that political intervention alters not merely the speed but the character of script change. Under natural transmission, change is distributed relatively evenly across feature domains. Under political intervention, change concentrates in features that define a script's fundamental architecture: pictographic origin, modular composition, and logographic component showed the highest intervention-to-natural rate ratios among all 50 features, with time-weighted point estimates exceeding 50$\times$ (bootstrap 95\% CIs are wide, reflecting the small natural-transmission rates in the denominators: pictographic origin 112$\times$ [1.10, 898]; modular composition 92$\times$ [0.48, 787]; logographic component 51$\times$ [0.36, 517]). The width of these intervals underscores that the individual ratios should be interpreted as indicators of direction and magnitude rather than precise multipliers; the primary evidence for selective rewriting is the low profile correlation ($\rho = 0.320$) rather than any single ratio. Political power, in other words, does not merely accelerate the clock; it rewrites the mechanism.

Conversely, stroke-based construction, curved strokes, and boustrophedon directionality showed near-zero change rates under both natural and political conditions. The physical substrate of writing resists transformation even under political coercion --- a limit to the reach of state power into embodied practice.

\begin{figure}[H]
\centering
\includegraphics[width=0.9\textwidth]{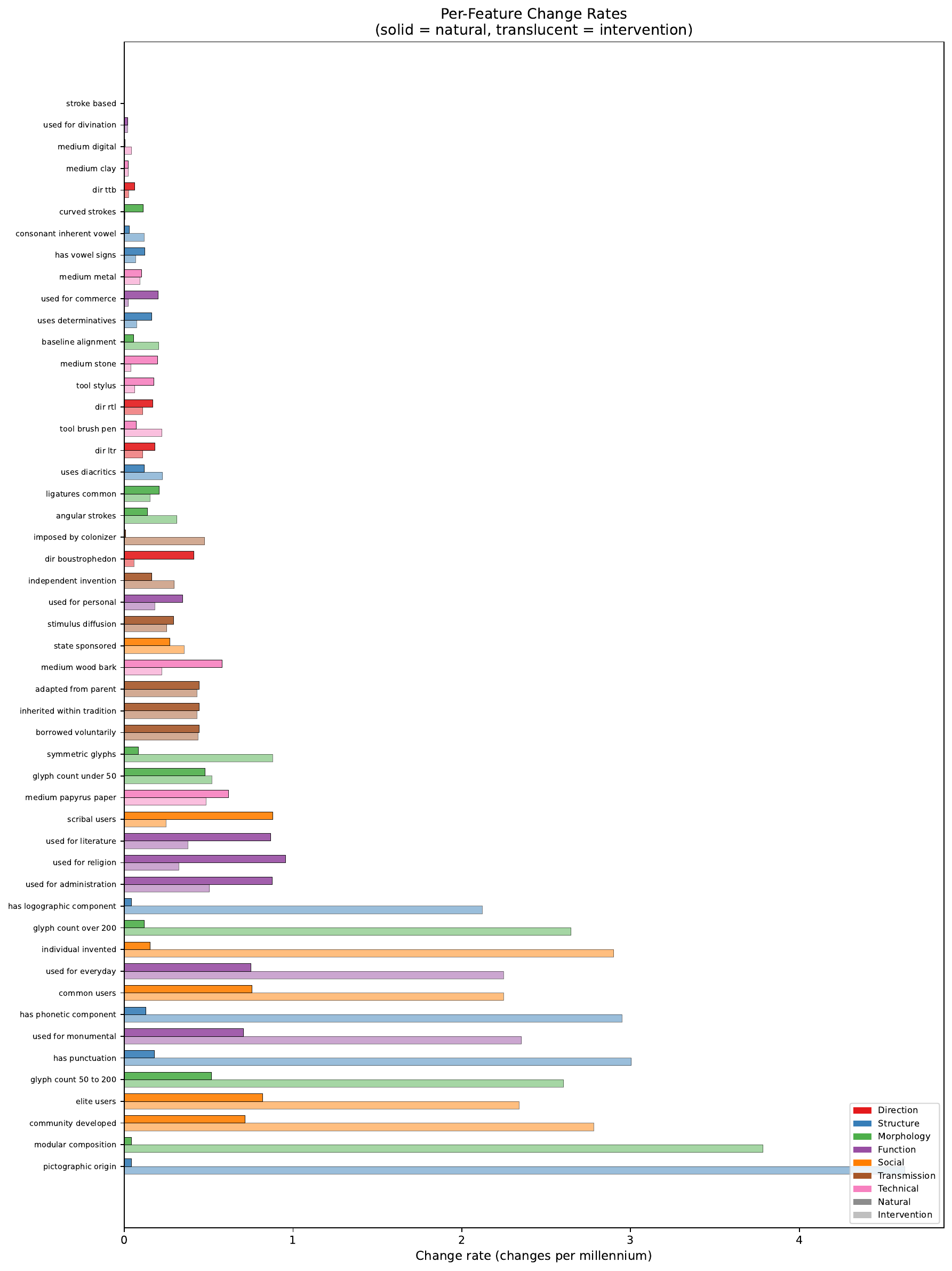}
\caption{Figure 4. Per-character substitution rates for 50 structural features, ranked from most conservative to most variable. Color indicates feature domain. Left bars: natural transmission; right bars: political intervention. Features tied to physical substrate (directionality, medium, stroke construction) change at rates below 0.05/millennium; social and structural features exceed 1.0/millennium.}
\label{fig:4}
\end{figure}

\subsection{3.4 Political Intervention Breaks the Clock}

The association between political intervention intensity and deviation from the molecular clock was substantial and highly significant (Spearman $\rho = 0.556$, $p < 10^{-4}$). Scripts that underwent political intervention deviated from the clock by a median of 1,169 years, compared with 879 years for naturally transmitted scripts.

The Kruskal-Wallis test confirmed that deviation from the clock differs significantly across intervention types (H = 25.5, $p < 0.0001$). Pairwise Dunn's post-hoc tests showed that "none" differed significantly from "stimulus diffusion," "colonial imposition," and "reform."

To test whether this pattern is driven by stimulus diffusion --- a category whose distance from parent scripts is partly definitional --- we repeated the analysis excluding all stimulus diffusion scripts (n = 71). The Spearman correlation decreased but remained significant ($\rho = 0.257$, $p = 5.9 \times 10^{-4}$), and the Kruskal-Wallis test on deviation by type remained significant (H = 16.6, $p = 0.0023$). The association between political intervention and clock deviation is robust to the exclusion of stimulus diffusion.

\subsection{3.5 The Ceiling Effect}

Writing was independently invented only four times in documented history --- in Mesopotamia, Egypt, China, and Mesoamerica --- always in environments where no existing writing system was present. To test whether existing writing systems suppress the emergence of new ones, we used two complementary approaches.

First, a Fisher's exact test compared the frequency of independent invention in "ceiling" environments (where a writing system was already in use) versus "non-ceiling" environments. The odds ratio was 0.054 ($p < 10^{-6}$), indicating that the presence of an existing script reduces the probability of independent invention by more than 18-fold. We acknowledge that this test is sensitive to the assumed number of independent inventions; if Egyptian hieroglyphs are reclassified as stimulus diffusion from Mesopotamia (reducing the count to 3), the odds ratio decreases further (OR = 0.036), strengthening rather than weakening the ceiling effect.

Second, a Cox proportional hazards model was fitted to the survival data of all 297 scripts (excluding 3 with incomplete data), with ceiling status as the primary covariate and script type, function, user class, political context, symbol count, and full-writing-system status as controls. Scripts originating under a ceiling --- in environments where another writing system was already present --- had a hazard ratio of 1.99 for extinction ($p = 0.030$, 95\% CI: 1.07--3.71), meaning they face approximately twice the extinction risk of scripts that emerged in writing-free environments.

The Ryukyu Islands provide an illustrative case study for the ceiling effect. Four parallel notation systems developed independently in Okinawa --- Sūchūma (base-5 numerical notation, ~13th century), Kaidā glyphs (pictographic notation, ~17th century; Rosa 2010, 2016), Barazan (knotted straw calculation), and Dāhan (household emblems) --- all under the "ceiling" of Japanese kana. Despite centuries of use, none of the four evolved phonetic encoding or expanded beyond its original functional domain. All four were extinguished following the Meiji government's assimilation of the Ryukyu Kingdom in 1879. We note that alternative explanations --- small population size, limited functional demand, or social stratification of literacy --- cannot be excluded for any individual case; the statistical evidence for the ceiling effect rests on the cross-system analysis (n = 297), not on the Ryukyu case alone.

\textbf{Colonial contact as a confound.} To test whether the ceiling effect is independent of colonial contact, we added colonial contact as a binary covariate to the Cox model. Colonial contact itself was a strong independent predictor of extinction (HR = 5.25, 95\% CI: 2.03--13.60, p = 0.0006). With colonial contact in the model, the ceiling effect decreased from HR = 1.99 (p = 0.030) to HR = 1.81 (p = 0.064), falling below conventional significance. The AIC of the extended model (1,270.5) was lower than the original (1,277.1), indicating improved fit. This result suggests that the ceiling effect and colonial contact are partially confounded --- scripts born under a ceiling are disproportionately likely to encounter colonial powers --- and that the independent contribution of ceiling status to extinction risk, while substantively meaningful (HR = 1.81), cannot be distinguished from colonial contact with the current sample size. The Fisher's exact test for independent invention (OR = 0.054) is not affected by this confound, as it concerns the suppression of invention, not extinction. We therefore maintain the ceiling effect claim for the suppression of independent invention, while noting that its extension to extinction risk requires further investigation.

\subsection{3.6 The Destruction of Writing Systems}

\subsubsection{3.6.1 Individual Script Replacement Events}

We identified and scored 30 major script replacement events (Figure 5). The destruction score integrates five equally-weighted normalized components (see Methods 2.6). To assess robustness to the choice of equal weighting, we performed a sensitivity analysis in which component weights were drawn from a Dirichlet distribution across 1,000 random configurations. The overall rank order was highly stable (Spearman $\rho = 0.989$ between equal-weight ranks and median ranks across configurations, $p < 10^{-24}$).

The top five by integrated destruction score are:

\begin{enumerate}
  \item \textbf{Arabic $\rightarrow$ Cyrillic, Uzbek (1940)}: 0.777 (median rank: 2, 95\% rank CI: [1, 9])
  \item \textbf{Arabic $\rightarrow$ Latin, Turkish (1928)}: 0.764 (median rank: 4, 95\% rank CI: [1, 14])
  \item \textbf{Arabic $\rightarrow$ Cyrillic, Kazakh (1940)}: 0.762 (median rank: 4, 95\% rank CI: [2, 10])
  \item \textbf{Arabic $\rightarrow$ Cyrillic, Tajik (1939)}: 0.751 (median rank: 5, 95\% rank CI: [3, 13])
  \item \textbf{Arabic $\rightarrow$ Cyrillic, Turkmen (1940)}: 0.742 (median rank: 7, 95\% rank CI: [4, 15])
\end{enumerate}

The dominance of Soviet Cyrillicization events reflects the scale, speed, and completeness of these transitions: entire writing systems were replaced within a single decade across Central Asia. The Turkish alphabet reform of 1928 ranks second by equal weighting but has a wider rank CI [1, 14], indicating that its position is more sensitive to the relative importance assigned to transition speed versus other components.

The Spanish destruction of Maya script (equal-weight rank 7, score = 0.697, median rank 8, 95\% rank CI: [1, 18]) is qualitatively distinct from all other events: it eliminated one of only four independent inventions of writing in human history. Its wide rank CI reflects a structural property of the scoring system: when weight shifts toward feature distance (where Maya scores maximally), its rank rises to 1; when weight shifts toward transition speed (where it scores low due to the centuries-long pace of cultural replacement), its rank falls. The irreplaceable loss of an independently evolved writing tradition represents a dimension of destruction that no scalar score can fully capture, and we flag this as a fundamental limitation of quantitative destruction metrics.

The Japanese elimination of Ryukyuan scripts ranks 12th (Kaidā glyphs $\rightarrow$ Hiragana, 1879, score = 0.591). This finding is presented with the awareness that the author is a Japanese national, and that the destruction of Ryukyuan writing systems by the Empire of Japan constitutes a case of imperial script elimination within the author's own national history.

\begin{figure}[H]
\centering
\includegraphics[width=0.9\textwidth]{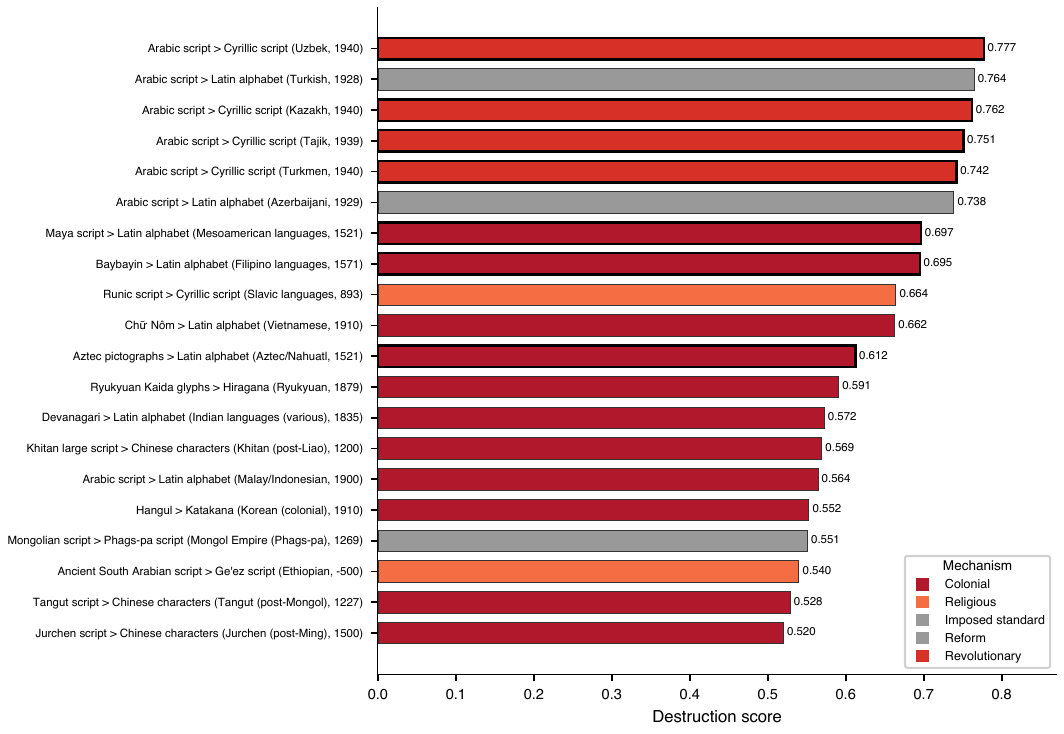}
\caption{Figure 5. Top 20 script replacement events ranked by integrated destruction score. Color indicates mechanism. Sensitivity analysis confirms rank-order stability (Spearman $\rho = 0.989$ across 1,000 weight configurations).}
\label{fig:5}
\end{figure}

\subsubsection{3.6.2 Imperial Script Destruction Rankings}

Attributing each script extinction to the political entity responsible yields the following ranking (Table 1):

\begin{table}[H]
\centering
\caption{Imperial Script Destruction Rankings}
\label{tab:1}
\begin{tabular}{lcccl}
\toprule
Empire & Scripts eliminated & Scripts contacted & Extinction rate & Notable victims \\
\midrule
Spanish Empire & 6 & 12 & 50.0\% & Maya, Baybayin, Isthmian, Mixtec-Puebla, Aztec pictographs, Testerian \\
Empire of Japan & 3 & 9 & 33.3\% & Kaidā glyphs, Sūchūma, Barazan \\
French Empire & 2 & 15 & 13.3\% & Ch\~{u} Nôm, Mi'kmaw hieroglyphs \\
Chinese Empires (Jin/Ming/Qing) & 3 & 26 & 11.5\% & Jurchen, Khitan Large, Khitan Small \\
\bottomrule
\end{tabular}
\end{table}

The addition of contact denominators reshapes the ranking. The Spanish Empire retains the highest extinction rate (50\%), consistent with its absolute count. However, the Empire of Japan rises to second place (33.3\%), surpassing the Chinese Empires (11.5\%) despite an equal absolute count. This reflects the concentrated destruction of Ryukyuan notation systems --- three of nine contacted scripts eliminated within a single generation. The Chinese Empires, by contrast, coexisted with 26 scripts over two millennia, absorbing only three through successive dynastic consolidations. The French Empire, despite administering vast colonial territories, directly eliminated only two scripts (13.3\%), suggesting that French colonial policy was less directly destructive to indigenous writing systems than Spanish or Japanese policy, even if its broader linguistic effects were substantial.

\subsubsection{3.6.3 Script Diversity Through Time}

The diversity timeseries (Figure 6) reveals a long, slow increase in the number of living scripts from approximately 4 systems at $-$5000 BCE to 100 systems by 1500 CE, followed by a rapid increase to a peak of 146 systems by 2000 CE. However, the colonial era (1500--1900 CE) imposed a severe disruption: the script extinction rate during this period was 8.75 extinctions per century, compared to a background rate of 2.07 extinctions per century --- a 4.2-fold increase.

The paradoxical increase in total script diversity \textit{despite} elevated colonial extinction is explained by an even higher rate of script creation during the same period, largely driven by stimulus diffusion: exposure to European writing prompted the creation of dozens of new scripts across Africa, Southeast Asia, and the Pacific. This pattern is analogous to an adaptive radiation following a mass extinction in biological evolution --- the ecological niches vacated by extinguished scripts were filled by newly created ones, though the newly created scripts are structurally less diverse than those they replaced.

\begin{figure}[H]
\centering
\includegraphics[width=0.9\textwidth]{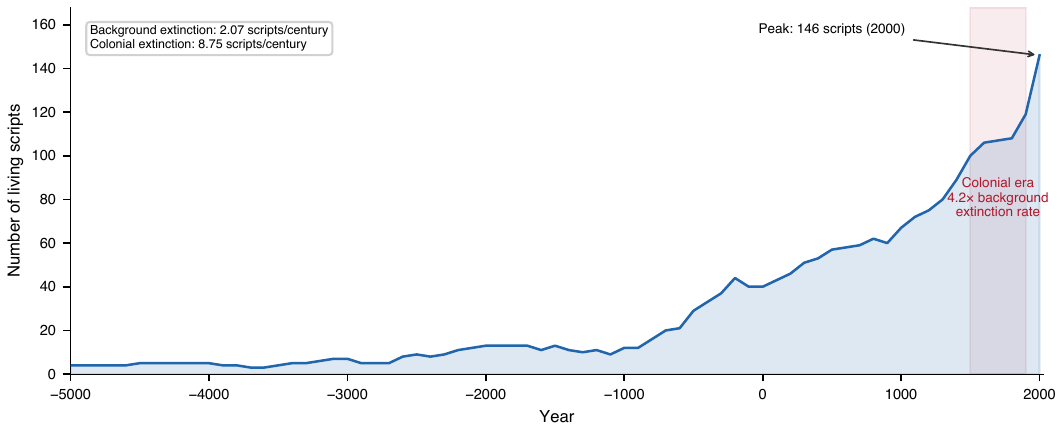}
\caption{Figure 6. Number of living writing systems through time ($-$5000 BCE to 2000 CE). Shaded red: colonial era (1500--1900 CE). Colonial extinction rate = 8.75 scripts/century (4.2$\times$ background rate). Peak: 146 scripts (2000 CE).}
\label{fig:6}
\end{figure}

\newpage

\section{4. Discussion}

\subsection{4.1 The Clock, Its Mechanism, and Its Uncertainty}

The central finding of this study is that writing systems evolve with a detectable regularity --- a molecular clock --- and that political power breaks it. The Mk+$\Gamma$ strict clock outperforms all alternative models ($\Delta\text{BIC} = -4.1$ versus Mk+$\Gamma$ relaxed; $\Delta\text{BIC} = -1{,}364.7$ versus Mk strict), and the estimated substitution rate of 0.226 characters per character per millennium provides a baseline against which deviations can be measured.

The introduction of Gamma-distributed rate variation across characters resolved a concern raised by the simpler Mk analysis. Under Mk, the strict clock was favored with $\Delta\text{BIC} = -14.4$, but this preference could have reflected model misspecification: if characters evolve at very different rates, the relaxed clock's branch-level rate variation parameters may be absorbed by unmodeled character-level variation, causing the strict clock to win by default. The Mk+$\Gamma$ model addresses this directly. Its shape parameter $\alpha = 0.51$ (95\% CI: 0.46--0.58) confirms that character-level rate variation is extreme --- most characters change slowly while a few evolve rapidly --- and the $\Delta\text{BIC}$ of 1,364.7 shows that ignoring this variation severely degrades model fit. That the strict clock remains favored even after this correction provides substantially stronger evidence for clock-like evolution than the Mk analysis alone.

The per-character rate analysis offers a first empirical answer to the question of why writing systems might evolve at a regular pace. The 50 features span more than three orders of magnitude in substitution rate, from stroke-based construction (0.001/millennium) to pictographic origin (2.17/millennium). This variation is not random: it follows a hierarchy of functional constraint. Features tied to the physical substrate of writing --- directionality, writing medium, stroke construction --- change at rates below 0.05/millennium, constrained by motor habits, material properties, and the embodied practices of scribal communities. Features describing a script's social positioning or high-level structural architecture --- pictographic origin, modular composition, user class --- change at rates exceeding 1.0/millennium, responding to shifts in political context, functional demand, and cultural contact.

The aggregate clock emerges, we suggest, because the conservative features act as a flywheel. While social and structural features fluctuate with every political transition, the physical substrate changes only on the timescale of technological revolutions (the shift from clay to papyrus, from brush to movable type). The overall rate $q$ is a weighted average of fast and slow features, and its regularity reflects the dominance of the slow features in the mixture --- a pattern captured by the low $\alpha$ of the Gamma distribution, which ensures that the many slowly evolving features anchor the mean.

This remains, however, a post hoc account rather than a theoretical prediction. We do not yet have a formal model analogous to the neutral theory's prediction of constant substitution rates in biology (Kimura 1968). The comparison with Lieberman et al.'s (2007) finding that English verbs regularize at a rate proportional to the inverse square root of their usage frequency is instructive: in their case, a population-genetic mechanism (frequency-dependent selection) provides the theoretical foundation. In ours, the mechanism --- functional constraint on physical features --- is plausible but not yet formalized. We identify the development of a mechanistic theory of script change rates as a priority for future work.

The 95\% confidence interval of the rate estimate (0.034--1.22 under Mk+$\Gamma$) spans approximately 36-fold. This width, combined with the MAE of 879--1,110 years, means the clock cannot be used for precise dating and should not be compared to the precision of molecular clocks in biology. We interpret the clock not as a chronometer but as a diagnostic instrument. Its value lies in detecting systematic deviations from baseline evolutionary rates, and these deviations are robust: the Spearman correlation between intervention intensity and clock deviation ($\rho = 0.556$) survives the exclusion of stimulus diffusion events ($\rho = 0.257$, $p = 5.9 \times 10^{-4}$), the exclusion of 9 politically contaminated features ($\rho = 0.505$, $p < 0.0001$), and the transition from Mk to Mk+$\Gamma$.

We note that the BIC difference between Mk+$\Gamma$ strict and Mk+$\Gamma$ relaxed ($\Delta\text{BIC}$ = 4.1) constitutes positive but not strong evidence on the Kass and Raftery (1995) scale. The overwhelming evidence is for character-level rate variation ($\Delta\text{BIC}$ = 1,364.7 for Mk+$\Gamma$ versus Mk); the choice between strict and relaxed branch-level clocks is more finely balanced. A larger dataset --- or the inclusion of scripts from underrepresented regions --- could shift this balance toward the relaxed clock. The present finding should be read as: a strict clock is the best-fitting model for N = 300, not as a claim that branch-level rate variation is absent.

\subsection{4.2 How Power Breaks the Clock}

The deviation analysis shows that political intervention increases clock deviation (MAE: 1,110 versus 879 years). But the per-character rate analysis reveals something more specific: political intervention does not merely accelerate the clock. It changes which features move.

Under natural transmission, change is distributed relatively evenly across feature domains, and the rate profiles of naturally transmitted and politically mediated scripts are only weakly correlated (Spearman $\rho = 0.320$). Under political intervention, change concentrates in the features that define a script's fundamental architecture --- pictographic origin, modular composition, logographic component (see Results 3.3 for per-feature rate ratios with bootstrap confidence intervals) --- precisely the features that distinguish an alphabet from a logosyllabary, a featural script from a syllabary. The low correlation between natural and intervention rate profiles ($\rho = 0.320$) confirms that this is not a uniform acceleration but a qualitative shift in which features are subject to change. Political reform --- Atatürk's replacement of Arabic with Latin, the Soviet Cyrillicization of Central Asia --- does not merely speed up the evolutionary clock. It reaches into the mechanism and replaces the gears.

Conversely, the most conservative features --- stroke construction, boustrophedon directionality, writing medium --- resist change even under political coercion. No decree can instantly retrain the muscle memory of a scribal population, or the material infrastructure of a writing culture. The physical substrate of writing represents a limit to the reach of state power into embodied practice.

This pattern --- deep structural replacement under political pressure, physical continuity in the substrate --- has an analogy in biology. Horizontal gene transfer in bacteria can replace metabolic pathways while leaving the ribosomal machinery intact. The comparison is imperfect, but the structural parallel is suggestive: in both cases, the most functionally constrained components resist replacement even when the less constrained components are freely exchanged.

\subsection{4.3 The Ceiling Effect and the Ecology of Writing}

Writing was independently invented at most four times --- in Mesopotamia, Egypt, China, and Mesoamerica --- and each time in an environment where no prior writing system existed. The ceiling effect (Fisher's exact OR = 0.054, $p < 10^{-6}$) suggests that existing writing systems suppress independent invention, analogous to competitive exclusion in ecology.

We acknowledge that this odds ratio could be seen as restating a historical truism: civilizations advanced enough to independently invent writing tend to exist in regions where writing already spread. The ceiling effect claim is stronger than this truism in one specific respect: the Cox survival analysis shows that ceiling status predicts extinction risk (HR = 1.99, p = 0.030), not merely failure to invent. Scripts born under a ceiling are not merely prevented from evolving into full writing --- they are also more likely to die.

However, a robustness check qualifies this extension. When colonial contact is added as a covariate, the ceiling effect decreases to HR = 1.81 (p = 0.064), while colonial contact itself emerges as a powerful independent predictor of extinction (HR = 5.25, p = 0.0006). The partial confound is expected: scripts born under a ceiling are disproportionately likely to encounter colonial powers, because both conditions --- the presence of existing writing and the presence of colonizers --- correlate with geographic proximity to literate, expansionist civilizations.

We therefore maintain the ceiling effect claim for the suppression of independent invention, where the statistical evidence is unambiguous (OR = 0.054). Its extension to extinction risk is substantively plausible (HR = 1.81 remains a non-trivial effect size) but cannot be cleanly separated from colonial contact with the current sample. The Fisher's exact test is not affected by the colonial confound, as it concerns the conditions under which writing emerges, not the conditions under which it dies.

The Ryukyu Islands illustrate both the ceiling effect and its entanglement with colonial power. Four parallel notation systems --- Sūchūma, Kaidā glyphs, Barazan, and Dāhan --- developed under the ceiling of Japanese kana. None evolved phonetic encoding. All four were extinguished following the Meiji government's assimilation of the Ryukyu Kingdom in 1879. Was their failure to evolve caused by the ceiling, or by the colonial absorption that terminated them? The honest answer is that the two cannot be disentangled in a single case. The statistical evidence for the ceiling effect rests on the cross-system analysis (n = 297), not on any individual case study.

\subsection{4.4 Imperial Destruction and the Irreplaceable}

The destruction score ranking is dominated by Soviet Cyrillicization events (top 5 positions). Sensitivity analysis confirms that this dominance is robust to weight perturbation ($\rho = 0.989$). However, individual ranks are less stable: the Turkish reform has a 95\% rank CI of [1, 14], and Maya has the widest CI [1, 18] of any top-10 event.

A key distinction emerges from the per-contact extinction rates. The Spanish Empire extinguished 6 of 12 contacted scripts (50\%), the Empire of Japan 3 of 9 (33.3\%), the French Empire 2 of 15 (13.3\%), and the Chinese Empires 3 of 26 (11.5\%). The absolute-count ranking obscures the concentration of Japanese imperial destruction: three of nine contacted scripts eliminated within a single generation, targeting the full set of indigenous Ryukyuan notation systems. The Chinese Empires, by contrast, coexisted with 26 scripts over two millennia, absorbing only three through dynastic consolidation.

The Soviet Union is absent from the imperial destruction table despite dominating the individual event rankings. The reason is definitional: the table counts scripts eliminated --- writing systems whose last user community was forcibly transitioned. Soviet Cyrillicization replaced the use of Arabic script in Central Asia but did not eliminate Arabic script itself. By contrast, the Spanish destruction of Maya eliminated a script with no other user community. This criterion captures a real distinction --- the permanence of loss --- but underweights the scale and violence of replacements that leave the script surviving elsewhere.

The Spanish destruction of Maya (rank 7, score 0.697, rank CI [1, 18]) is qualitatively distinct from all other events. It eliminated one of only four independent inventions of writing --- an entire evolutionary lineage, not merely a single script. No scalar score can rank this loss against the speed and efficiency of Soviet Cyrillicization. We suggest that a separate typology distinguishing "branch extinction" (the loss of one script within a lineage) from "lineage extinction" (the loss of an independently evolved writing tradition) may be needed. By this typology, the Spanish destruction of Maya is the single most destructive event in the recorded history of writing.

\subsection{4.5 Positionality}

The author is a Japanese national. The finding that the Empire of Japan extinguished three writing systems in the Ryukyu Islands --- ranking second in per-contact extinction rate at 33.3\% --- is presented as quantitative evidence, not as apology. That a Japanese researcher independently arrives at this finding through computational analysis, without external pressure to do so, may serve as a demonstration that quantitative methods can discipline the self-serving tendencies of national historiography.

\subsection{4.6 Limitations}

We identify seven limitations that affect the interpretation of these findings.

First, feature coding circularity. The character matrix was coded by a large language model (Claude, Anthropic) whose training data includes information about the archaeological dates of writing systems. Four considerations mitigate this concern: (a) inter-rater reliability testing with two independent human coders blind to script dates and political context, on a stratified sample of 40 scripts (n = 2,000 binary judgments), showed high agreement (Cohen's $\kappa = 0.877$, human-LLM $\kappa = 0.929$, Fleiss' $\kappa = 0.911$), (b) per-domain $\kappa$ ranged from 0.84 to 0.91, with only one feature exhibiting low $\kappa$ due to prevalence imbalance (curved\_strokes: $\kappa = -0.04$, raw agreement 92.9\%), (c) the clock signal survives the exclusion of 9 features conceptually linked to political context (41-feature analysis: Spearman $\rho = 0.505$, $\Delta\text{BIC} = -15.2$), and (d) the clock fails systematically where political interventions distort the feature record, a pattern inconsistent with simple data leakage. While a 40-script stratified sample provides substantially stronger validation than prior pilot testing, the IRR coders were trained psychologists rather than expert grammatologists; full-scale blind recoding by specialists in writing systems remains desirable for journal submission.

Second, the 95\% CI of the clock rate spans 36-fold (0.034--1.22 under Mk+$\Gamma$). The clock is diagnostic, not chronometric.

Third, the Mk+$\Gamma$ model assumes Gamma-distributed rate variation. While this is a standard assumption in molecular phylogenetics, the true distribution of character-level rate variation in writing systems is unknown. The strong fit ($\Delta\text{BIC} = -1{,}364.7$) suggests that the Gamma approximation captures the dominant pattern, but alternative rate distributions have not been tested.

Fourth, small sample sizes in several intervention categories limit statistical power. Colonial imposition (n = 7), reform (n = 12), and imposed standard (n = 9) are most affected.

Fifth, the destruction score uses equal weighting of five components without theoretical justification. Sensitivity analysis shows that the overall rank order is robust ($\rho = 0.989$), but individual event ranks can shift substantially.

Sixth, the ceiling effect's extension to extinction risk is confounded with colonial contact. The Fisher's exact test for suppression of invention (OR = 0.054) is robust, but the Cox model's ceiling HR decreases from 1.99 to 1.81 (p = 0.064) when colonial contact is controlled.

Seventh, the phylogenetic edges. Of 259 parent-child relationships, 79 (30.5\%) are established in major reference works, 174 (67.2\%) reflect scholarly consensus, and 6 (2.3\%) are inferred. The evidence classification was performed by the LLM and has not been independently verified.

\subsection{4.7 Future Directions}

Three extensions are planned. First, the development of a mechanistic theory of script change rates --- formalizing the relationship between functional constraint, feature-level substitution rates, and aggregate clock behavior. The per-character rate data presented here provide the empirical foundation; what is missing is a formal model analogous to neutral theory in population genetics. Second, the application of the methodology to undeciphered scripts, beginning with Linear A using the SigLA database. Third, the development of a formal "Red List" of endangered writing systems, analogous to the IUCN Red List for species, using the Cox survival model to estimate extinction probabilities for currently living scripts.

\section{5. Conclusion}

Writing systems are cultural replicators. Like biological organisms, they descend from ancestors, accumulate changes, diversify into distinct lineages, and go extinct. Unlike biological organisms, they can be created, transformed, or destroyed by political decree.

The molecular clock of writing, at $q = 0.226$ substitutions/character/millennium (Mk+$\Gamma$ strict), provides a quantitative baseline for measuring the pace of script evolution. The clock is imprecise --- its confidence interval spans 36-fold --- but its deviations are informative. Where the clock deviates most from expectation, political power has intervened. And the nature of that intervention is now visible at the level of individual features: political power selectively rewrites the deep structural architecture of scripts while leaving their physical substrate intact.

Four times in human history, a society invented writing from nothing. Every other writing system on Earth is a descendant, a copy, or an echo of those four acts. The ceiling effect ensures that independent invention cannot recur in any environment where writing already exists. What took millennia to create can be destroyed in a decade --- and the destruction is not equal. The Spanish Empire eliminated half of all writing systems it encountered. The Empire of Japan eliminated a third. These numbers are not accusations; they are measurements. The data presented here are an attempt to measure the distance between creation and destruction --- and to name the forces responsible.

\newpage

\section{Acknowledgments}

We thank Akiko Tamamura and Miki Maeda (Research Institute of Criminal Psychiatry / Sexual Offender Medical Center) for research assistance with inter-rater reliability coding.

\newpage

\section{Supplementary Materials}

\begin{figure}[H]
\centering
\includegraphics[width=0.9\textwidth]{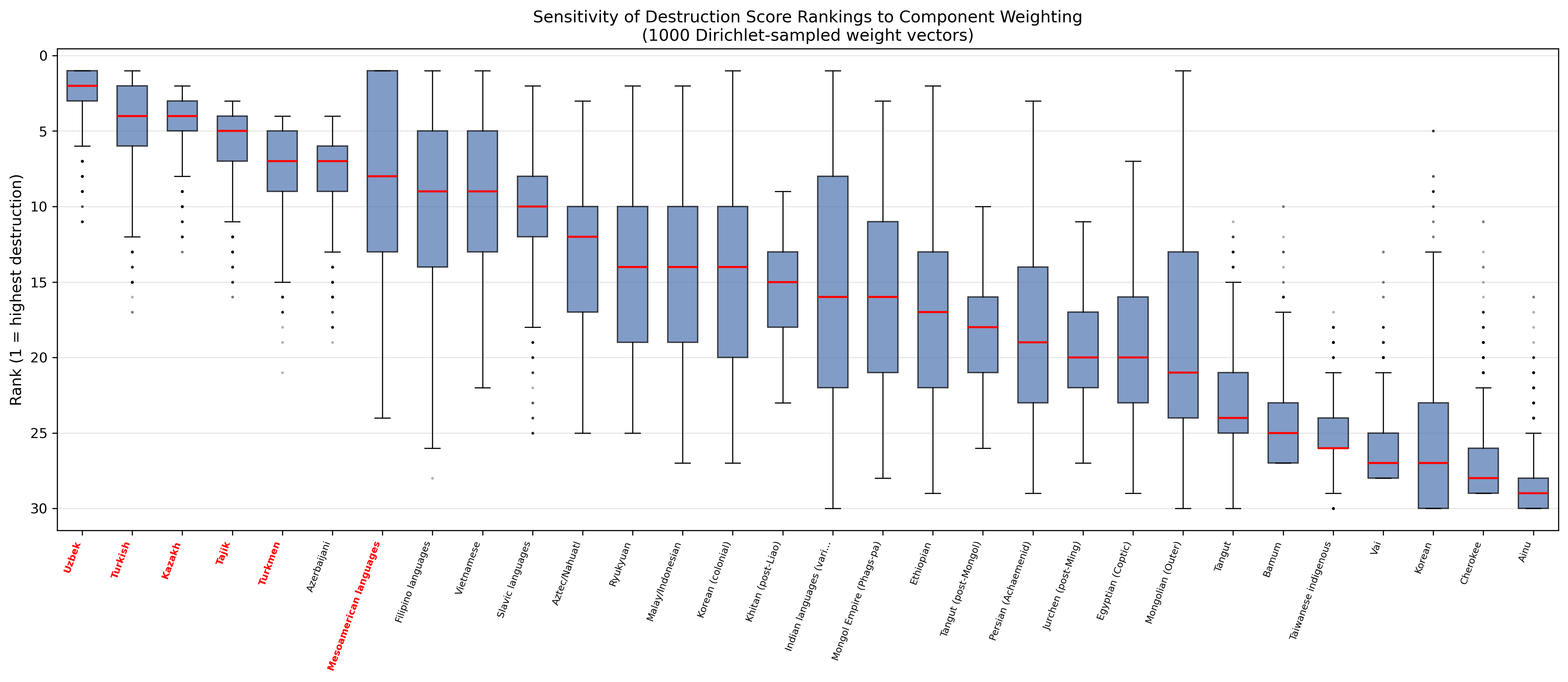}
\caption{Figure S1. Sensitivity analysis of destruction score rankings. Box plots show rank distribution across 1,000 random weight configurations (Dirichlet-sampled). Whiskers: 95\% rank CI.}
\label{fig:S1}
\end{figure}

\begin{figure}[H]
\centering
\includegraphics[width=0.9\textwidth]{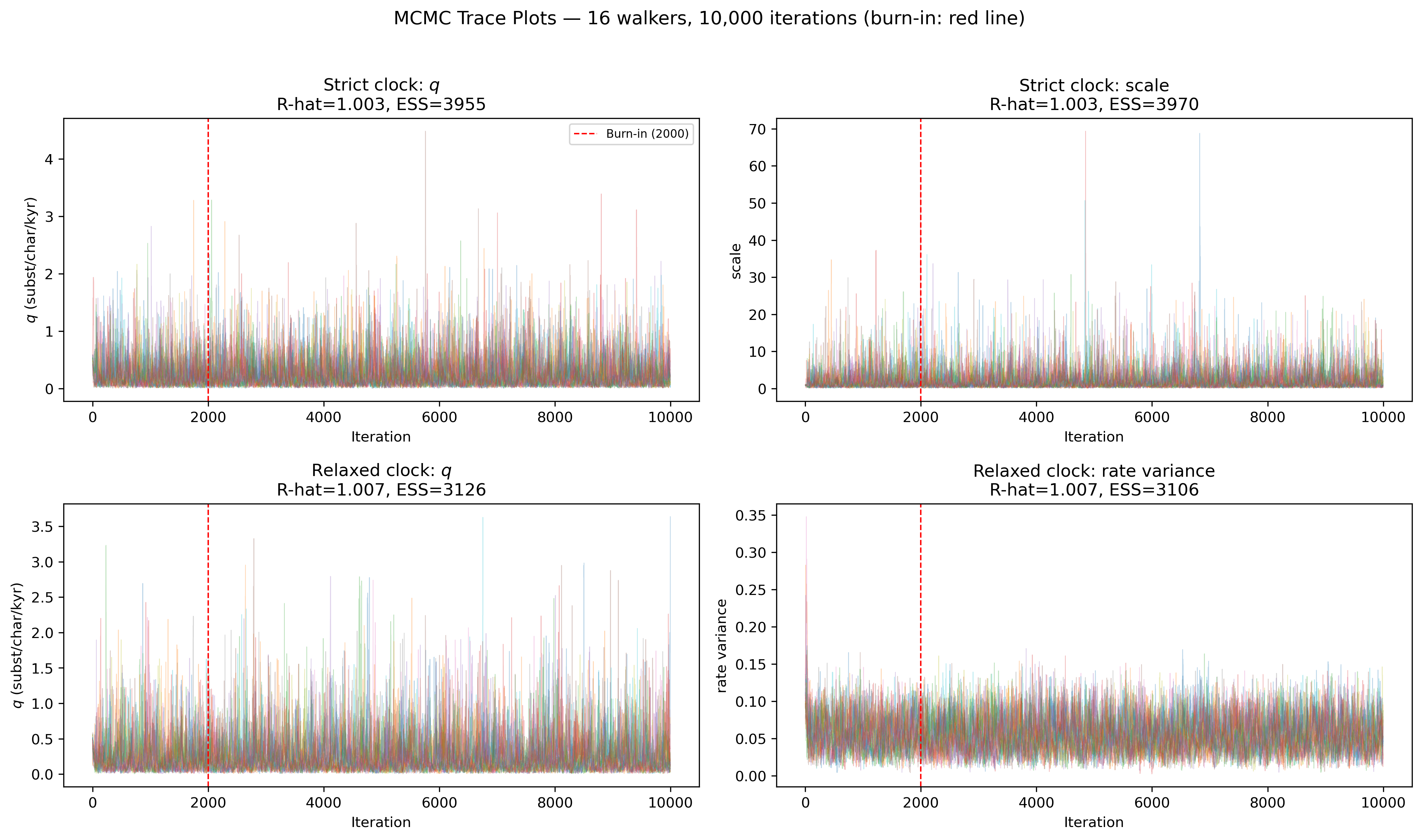}
\caption{Figure S2. MCMC trace plots for Bayesian clock analysis. Top: strict clock (log q and log scale). Bottom: relaxed clock (log q, log scale, log rate\_var). All 16 walkers shown. Gelman-Rubin $\hat{R} < 1.01$ for all parameters.}
\label{fig:S2}
\end{figure}

\begin{figure}[H]
\centering
\includegraphics[width=0.9\textwidth]{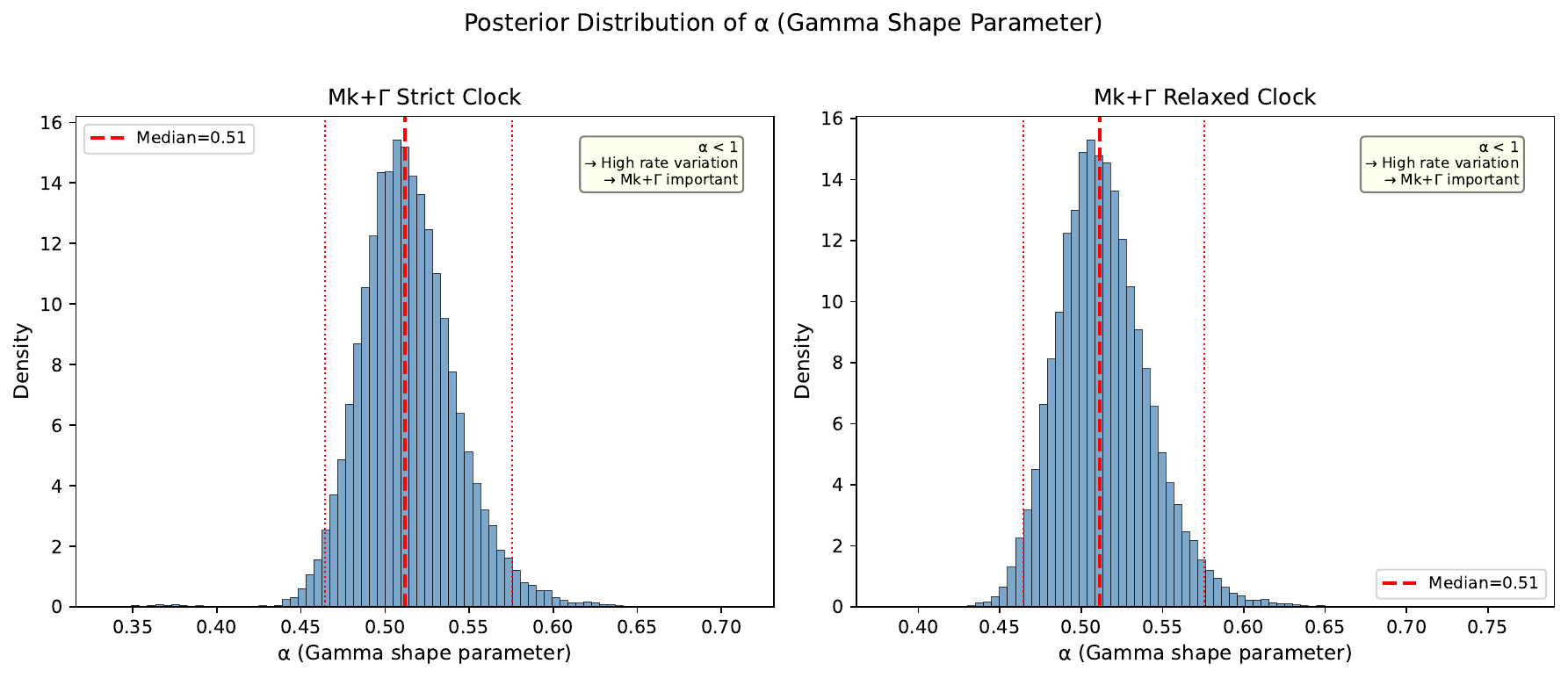}
\caption{Figure S3. Posterior distribution of the Gamma shape parameter $\alpha$ from the Mk+$\Gamma$ strict clock model. Median $\alpha = 0.51$ (95\% CI: 0.46--0.58). Values below 1 indicate strong rate heterogeneity among characters.}
\label{fig:S3}
\end{figure}

\begin{figure}[H]
\centering
\includegraphics[width=0.9\textwidth]{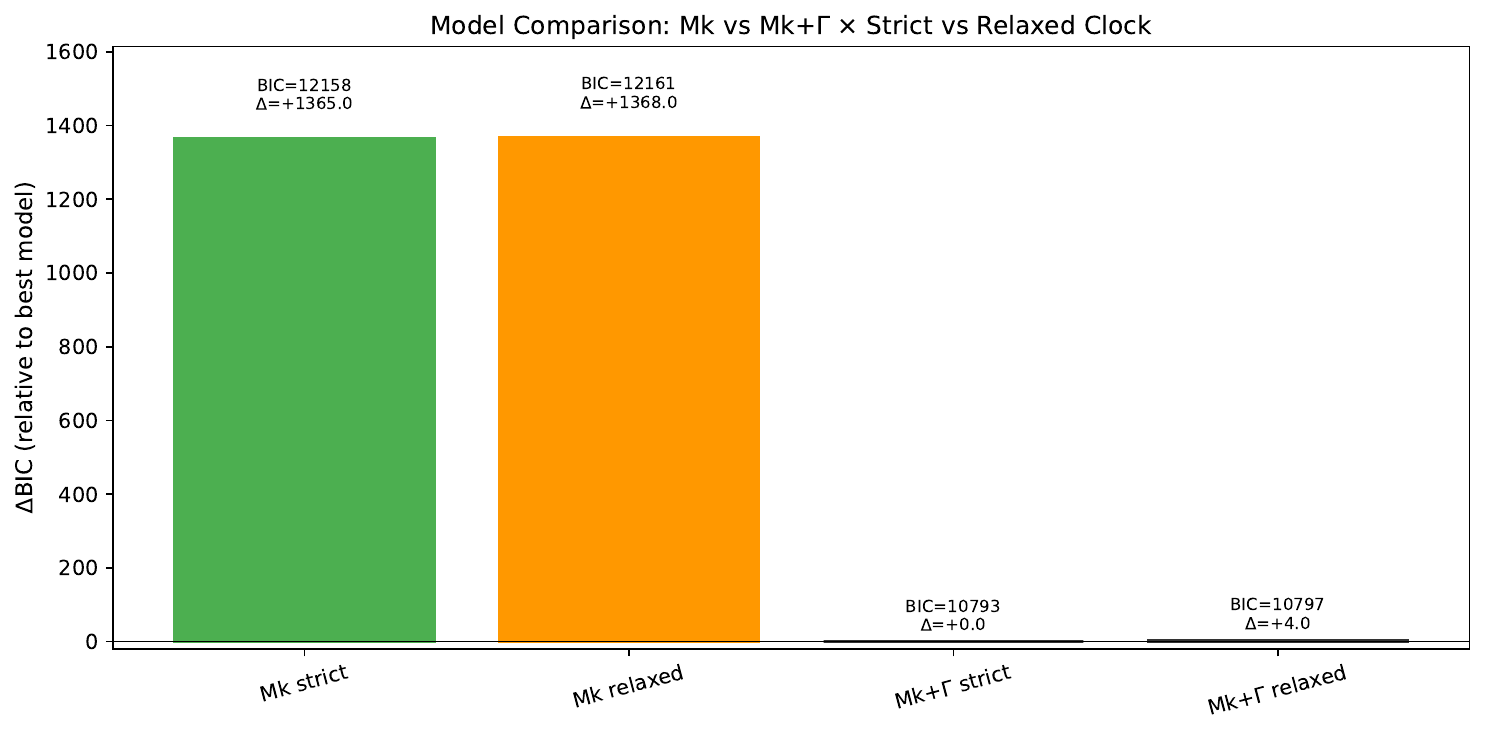}
\caption{Figure S4. BIC comparison across four models: Mk strict, Mk relaxed, Mk+$\Gamma$ strict, Mk+$\Gamma$ relaxed. Mk+$\Gamma$ strict is the best-fitting model (lowest BIC).}
\label{fig:S4}
\end{figure}


\begin{thebibliography}{99}

\bibitem{Bentz2026} Bentz, C., \& Dutkiewicz, E. (2026). Humans 40,000 y ago developed a system of conventional signs. \textit{Proceedings of the National Academy of Sciences}, 123(9), e2520385123.

\bibitem{Coulmas2003} Coulmas, F. (2003). \textit{Writing Systems: An Introduction to Their Linguistic Analysis}. Cambridge University Press.

\bibitem{Daniels1996} Daniels, P. T., \& Bright, W. (Eds.). (1996). \textit{The World's Writing Systems}. Oxford University Press.

\bibitem{Felsenstein1981} Felsenstein, J. (1981). Evolutionary trees from DNA sequences: A maximum likelihood approach. \textit{Journal of Molecular Evolution}, 17(6), 368--376.

\bibitem{Gray2003} Gray, R. D., \& Atkinson, Q. D. (2003). Language-tree divergence times support the Anatolian theory of Indo-European origin. \textit{Nature}, 426(6965), 435--439.

\bibitem{Greenhill2017} Greenhill, S. J., Wu, C.-H., Hua, X., Dunn, M., Levinson, S. C., \& Gray, R. D. (2017). Evolutionary dynamics of language systems. \textit{Proceedings of the National Academy of Sciences}, 114(42), E8822--E8829.

\bibitem{Hosszú2024} Hosszú, G. (2024). Validation of graph sequence clusters through multivariate analysis: application to Rovash scripts. \textit{npj Heritage Science}, 12, 110.

\bibitem{Kass1995} Kass, R. E., \& Raftery, A. E. (1995). Bayes factors. \textit{Journal of the American Statistical Association}, 90(430), 773--795.

\bibitem{Kimura1968} Kimura, M. (1968). Evolutionary rate at the molecular level. \textit{Nature}, 217(5129), 624--626.

\bibitem{Lewis2001} Lewis, P. O. (2001). A likelihood approach to estimating phylogeny from discrete morphological character data. \textit{Systematic Biology}, 50(6), 913--925.

\bibitem{Lieberman2007} Lieberman, E., et al. (2007). Quantifying the evolutionary dynamics of language. \textit{Nature}, 449(7163), 713--716.

\bibitem{Mace2005} Mace, R., \& Holden, C. J. (2005). A phylogenetic approach to cultural evolution. \textit{Trends in Ecology \& Evolution}, 20(3), 116--121.

\bibitem{Pagel2007} Pagel, M., Atkinson, Q. D., \& Meade, A. (2007). Frequency of word-use predicts rates of lexical evolution throughout Indo-European history. \textit{Nature}, 449(7163), 717--720.

\bibitem{Robinson2007} Robinson, A. (2007). \textit{The Story of Writing}. Thames \& Hudson.

\bibitem{Rosa2010} Rosa, M. C. (2010). Ryukyuan writing systems and notation. [Conference paper].

\bibitem{Rosa2016} Rosa, M. C. (2016). Kaidā glyphs in the digital humanities. In \textit{Proceedings of JADH 2016}.

\bibitem{Yang1994} Yang, Z. (1994). Maximum likelihood phylogenetic estimation from DNA sequences with variable rates over sites: approximate methods. \textit{Journal of Molecular Evolution}, 39(3), 306--314.

\bibitem{Zuckerkandl1965} Zuckerkandl, E., \& Pauling, L. (1965). Evolutionary divergence and convergence in proteins. In \textit{Evolving Genes and Proteins} (pp. 97--166). Academic Press.

\end{thebibliography}
\end{document}